\documentclass[]{aiaa-tc}

\usepackage{lettrine,amsmath}
\usepackage{iepc2024}
\usepackage{url, xcolor, bm}
\usepackage{soul}

\IEPCsubmissionnumber{659}%

\title{Full Helicon thruster modeling}

\author{ D. Iannarelli \thanks{Presenting author, PhD student, DIAEE La Sapienza University of Rome} \,\, and A. Ingenito \thanks{Professor, School of Aerospace Engineering, La Sapienza University of Rome}
\\
{\normalsize\it{School of Aerospace Engineering, University ``Sapienza'', Via Salaria
851-881, Rome 00138, Italy}}\\
\\
F. Napoli \thanks{Researcher, NUC-PLAS division}\,\, ,
F. Cichocki \thanks{Researcher, NUC-PLAS division}\,\, ,
C. Castaldo \thanks{Researcher, NUC-PLAS division}\,\, ,
A. De Ninno \thanks{Researcher, NUC-PLAS division}\,\, ,
S. Mannori \thanks{Researcher, NUC-PLAS division} \\
{\normalsize\it{ENEA Frascati Research Center, Via Enrico Fermi 45, Frascati, Rome, Italy}}\\
\\
A. Cardinali \thanks{Researcher, associated to CNR}\\
{\normalsize\it{INAF, Istituto di Astrofisica e Planetologia Spaziali, via del Fosso del Cavaliere 100, 00133 Roma, Italy}} \\
{\normalsize\it{CNR, Istituto Sistemi Complessi, Politecnico di Torino, Corso Duca degli Abruzzi 24, 10129 Torino, Italy}} \\
\\
F. Taccogna \thanks{Researcher}
\\
{\normalsize\it{CNR, Istituto per la Scienza e Tecnologia dei Plasmi, Via Amendola, 122/D, 70126 Bari}}
}

\begin{document}

\maketitle

\begin{abstract}

This research work is part of a PhD project aiming to design a new helicon plasma thruster for cargo space missions, within a collaboration between the School of Aerospace Engineering of La Sapienza University and the ENEA Frascati research center. 
A 0-D global design model of the thruster is first established to identify the main operational conditions of the thruster, starting from a targeted performance in terms of specific impulse and thrust. These operational conditions are tested with dedicated simulations on the electromagnetic plasma-wave interaction within the ionization chamber, and on the electrostatic plasma acceleration within the magnetic nozzle. The former are carried out with a finite element electromagnetic code and confirm the technical feasibility of the identified operational RF power with a predefined chamber geometry. The latter are obtained with an electrostatic particle-in-cell code, which allows to estimate more precisely the achieved thruster performance. A sub-optimal thruster with an absorbed plasma power of around 1 kW, a thrust of around 10 mN and a specific impulse higher than 1000 s is successfully simulated. The established simulation framework will serve as the basis for the next optimization of the thruster geometry and operational conditions, with the ultimate goal of building and testing a new helicon thruster prototype.


\end{abstract}

\section*{Nomenclature}
\begin{center}
\begin{tabular}{lp{5in}}

$\dot{m}_\mathrm{i}, \dot{m}$  & = ion and total mass flow rates (kg/s)  \\
$m_\mathrm{i}, m_\mathrm{e}$    & = ion and electron elementary masses (kg) \\
$e$         & = elementary electronic charge (C) \\
$c, c_\mathrm{s}$         & = speeds of light and of plasma sound waves (m/s) \\
$C_\mathrm{r}, C_\mathrm{z}$ & = plasma density ratios along the radial and axial directions at the sheath edge \\
$\mu_{0}$         & = vacuum magnetic permeability (H/m) \\
$\epsilon_{0}$         & = vacuum dielectric constant (C$^{2}$/(Nm$^{2}$)) \\
$F$         & = thrust force (N) \\
$f_\mathrm{z,mag}$         & = axial magnetic force density (N/m$^3$) \\
$I_\mathrm{sp}$         & = specific impulse in time units (s) \\
$g_0$         & = standard gravity acceleration (m/s$^2$) \\
$\eta_{m}$      & = propellant utilization efficiency (-) \\
$\eta_{RF}$      & = RF power transmission efficiency to the plasma (-) \\
$\eta_{t}, \eta_\mathrm{div}$  & = total thruster efficiency and plume divergence efficiency (-) \\
$u_\mathrm{i}, u_\mathrm{z,i}$       & = ion fluid velocity and its axial component (m/s) \\
$P_\mathrm{RF}, $   & = input antenna power (W) \\
$P_\mathrm{wall}, P_\mathrm{ion}, P_\mathrm{exc}, P_{\Omega}$
& = power to wall, ionization and excitation powers, power absorbed from wave (W) \\
$r, z$  & = radial and axial coordinates ($m$) \\
$T_\mathrm{e}, T_\mathrm{e0}$ & = electron temperature (in energy units) and its peak value (J, eV) \\
$n_\mathrm{e}, n_\mathrm{i}, n_\mathrm{e0}$   & = electron and ion densities and their peak value (m$^{-3}$) \\
$\bar{n}_\mathrm{e}, \bar{n}_\mathrm{e,l}, n_\mathrm{e,f}$   & = plasma densities averaged over respectively the full chamber, the lateral chamber surface and the chamber exit surface (m$^{-3}$)\\
$A_\mathrm{f},A_\mathrm{l}$& = frontal and lateral surfaces of the ionization chamber ($m^{2}$)  \\
$L, R$  & = ionization chamber length and radius (m) \\
$\Delta V$  & = ionization chamber volume (m$^{3}$) \\
$\bar{n}_\mathrm{n}$  & = average neutral density inside the chamber (m$^{-3}$) \\
$R_\mathrm{ion},R_\mathrm{exc}$  & = ionization and excitation rates (m$^{3}$/s)  \\
$E_\mathrm{ion},E_\mathrm{exc}$  & = ionization and excitation threshold energies (J, eV)  \\
$\gamma_\mathrm{tot}, \gamma_\mathrm{tot,l}, \gamma_\mathrm{tot,f}$ & = energy transfer coefficient and its average over the lateral and front surfaces (-) \\
$\psi$  & = angle between the magnetic field line and the surface normal (rad, deg) \\
$k, k_{\parallel}, k_{\perp}$  & = total, parallel and perpendicular wave numbers (m$^{-1}$) \\
$\lambda$  & = parallel wave length (m) \\
$\omega$  & = wave angular frequency (rad/s) \\
$\omega_\mathrm{p,e},\omega_\mathrm{c,e}$  & = electron plasma and cyclotron frequencies (rad/s) \\
$a, L_\mathrm{a}$  & = antenna radius and length (m) \\
$B_{0}$  & = magnetostatic induction field (T, G) \\
$\bm{E}, \bm{B}$  & = Electric and magnetic induction field vectors (V/m) \\
$\bm{j}$  & = Electric current density vector (A/m$^2$)\\
$\bm{E}^{*}$  & = complex conjugate of the wave electric field vector (V/m) \\
$\phi, \phi_{\infty}, \phi_\mathrm{t}, \phi_\mathrm{exit}$  & = electrostatic potential, its value at infinity, at nozzle throat, and at electrons domain exit point (V) \\
$E_\mathrm{tot}$  & = total electron mechanical energy (J, eV) \\
$v_\mathrm{e}$  & = electron velocity (m/s) \\
$I_\mathrm{e \infty}, I_\mathrm{i \infty}$ & = electron and ion currents to infinity (A)\\
$C_{\infty}$ & = equivalent electric capacity (F) \\
$v_\mathrm{th,e}$ & = thermal electron velocity (m/s) \\
$\sigma_\mathrm{n}$ & = neutral cross section (m$^2$) \\
$Z_\mathrm{i}$ & = ion charge number (-)
\end{tabular}
\end{center}

\newpage

\section{Introduction}
The helicon thruster is a relatively new electric thruster in which the thrust is produced by accelerating, through
a magnetic nozzle, a propulsive plasma. This is obtained by coupling RF power with helicon waves inside a ionization
chamber (a dielectric tube where the propellant is injected) using a special type of antenna, known as \textit{helicon
antenna} \cite{chen91}, that is wrapped around the tube. The helicon thruster is an electrodeless thruster as it ionizes the
propellant without electrodes and requires no external cathode to neutralize the emitted plume, which is already
globally ambipolar. This simplifies the design and allows for a lighter thruster, not exposed to sputtering, and
therefore characterized by a longer lifetime. The ionization mechanism has already been tested experimentally
and helicon thrusters are designed to be mounted on small satellites with feeding powers from a few hundred W up to few kWs and thrust levels from few mN up to hundreds of mN.

The main goal of this research study is to model the entire thruster, simulating the plasma within both the helicon thruster ionization chamber and the magnetic nozzle, in order to estimate the overall propulsive performance of the device. 

First of all, a 0D global performance model is introduced in order to determine a first guess of sub-optimal operating conditions (electron temperature, mass flow, average plasma and neutral density), as a function of the target thruster performance (mainly thrust, specific impulse and total operating RF power).

Once the average plasma density and electron temperature in the source is known, the electromagnetic antenna-plasma interaction is modeled by a full-wave approach, with the goal of designing in realistic geometry the required  antenna to couple the correct power to the plasma.
In particular, the coupling of the helicon antenna with the plasma is solved using a 3D finite element method (FEM) electromagnetic code, accounting for a realistic geometry of both the antenna structure and the ionization chamber. 
The plasma is modeled as an homogeneous, anisotropic, equivalent dielectric with permittivity given by the Stix dielectric tensor \cite{STIX92} with collisional corrections. Electron-ion and electron-neutral collisions are considered in the plasma dielectric tensor using correction terms proportional to the collisional frequency \cite{card14,mela15}. This is equivalent to consider the plasma as a conductive, anisotropic, locally
homogeneous, collisional fluid, magnetized by an external magnetostatic field. The cold plasma approximation is considered adequate in describing the propagation of fast waves (helicons) in the cold plasma of the helicon thruster \cite{card14,mela15}. 
Although it has been shown that a non-uniform magnetic induction field might be relevant for an accurate estimation of the wave absorption \cite{card14}, in these full-wave simulations a constant and uniform axial field has been considered, given its negligible spatial variations within the ionization chamber (with a small aspect ratio $R/L$ as in the present case). 

Then, the plasma properties at the thruster exit section are considered by a particle-in-cell (PIC) code, called PICCOLO (PIC COde of LOw temperature plasma discharges), which simulates the entire magnetic expansion of the nozzle, thus estimating the produced thrust force. PICCOLO is based on the electrostatic PIC technique, in which ions and electrons are simulated as macro-particles subject to Newton’s equation and their charge is weighted at the nodes of a dedicated mesh (hence the name particle-in-cell to obtain the spatial distribution of electric charge density \cite{tacc23}). This is used by a Poisson’s equation solver to obtain the self-consistent electric field used to advance the charged particles into the next time step.

The simulation results obtained in this study (from both the full-wave and PIC codes) will be used in the future to better refine the parameters of the global performance model.
%
Therefore, this work will serve as the basis for the optimized design, manufacture and testing of a helicon thruster prototype. Future comparison of the experimental data with the numerical predictions will finally allow to further tune and validate the developed numerical models.

In the following, the 0-D performance model is presented in Sec.\,\ref{sec: 0D model}, the plasma-wave interaction model in Sec.\,\ref{sec: plasma wave model} and the PIC model for the magnetic nozzle in Sec.\,\ref{sec: PIC model}. The simulation results are discussed in Sec.\,\ref{sec: simulation results}, and finally conclusions are drawn in Sec.\,\ref{sec: conclusions}.


\section{0-D global design model}
\label{sec: 0D model}
In order to have a first-guess sizing of the Helicon plasma thruster, a 0-D global equilibrium model is used in order to compute operating parameters such as the  electron density $n_\mathrm{e0}$, the electron temperature $T_{e0}$ inside the source, and the total mass flow rate $\dot{m}$. This 0-D model assumes:

\begin{itemize}
\item A given thrust force $F$, as a target design parameter;
\item a given thruster specific impulse ($I_\mathrm{sp}$), as a target design parameter;
\item a given geometry of the ionization chamber, defined in terms of both its radius $R$ and length $L$. Such geometrical parameters may be changed in a iterative design for the global optimization of the thruster. However, this design procedure is not pursued here and is demanded to a successive work.
\end{itemize}

The total mass flow rate of the thruster can be directly estimated from the target parameters as:

\begin{equation}
 \dot{m} = \frac{F}{g_0 I_\mathrm{sp}} = \frac{F}{\eta_{m} u_\mathrm{z,i}},
 \label{eq: mass flow}
\end{equation}
where $I_\mathrm{sp}$ is the specific impulse in time units, $g_0$ is the standard gravity acceleration, $u_\mathrm{z,i}$ is the axial velocity of the ions (at infinity) and $\eta_{m} = \dot{m_\mathrm{i}}/\dot{m}$ is the propellant utilization efficiency. In this equation, we have assumed a negligible neutral exit velocity (compared to ions that are typically 10-20 times faster). 

From the energy conservation across the magnetic nozzle expansion, we can compute the final ion exit velocity from the predicted electric potential drop $\Delta \phi_\mathrm{MN}$. In fact, in a magnetic nozzle, the thermal electron energy is transformed into directed ion kinetic energy: 

\begin{align}
    \frac{1}{2} m_\mathrm{i} u_\mathrm{z,i}^{2} = & e \Delta \phi_\mathrm{MN},
    \label{eq: ion axial velocity}\\
    \Delta \phi_\mathrm{MN} = & \frac{T_\mathrm{e0}}{e} \left( -\ln\sqrt{\frac{2\pi m_\mathrm{e}}{m_\mathrm{i}}} + 0.5 \right) = \frac{C_\mathrm{shd} T_\mathrm{e0}}{e},
    \label{eq: delta phi drop}
\end{align}
where $T_\mathrm{e0}$ is the electron temperature (in energy units), $C_\mathrm{shd} \approx 5 $ is a non-dimensional constant depending on the electron to ion mass ratio, and the factor 0.5 models the ion acceleration in the pre-sheath inside the ionization chamber (where ions are still subsonic). Eq.\,\ref{eq: delta phi drop} represents the electric potential drop over a classical Debye sheath, which is actually a conservative approach, as shown in recent studies suggesting higher potential drops \cite{ahed20}. Combining Eqs.\,\ref{eq: ion axial velocity} and \ref{eq: delta phi drop}, and knowing that $g_0 I_\mathrm{sp} \approx \eta_{m}  u_\mathrm{z,i}$, it is possible to obtain a conservative estimate of the required electron temperature to achieve the target specific impulse:
\begin{equation}  
T_\mathrm{e0} = \frac{1}{2}\frac{m_\mathrm{i}}{\eta_\mathrm{m}^{2} C_\mathrm{shd}} (g_0 I_\mathrm{sp})^{2}.
\end{equation}

\begin{figure}[t!]
  \centering
  \includegraphics[width = \textwidth]{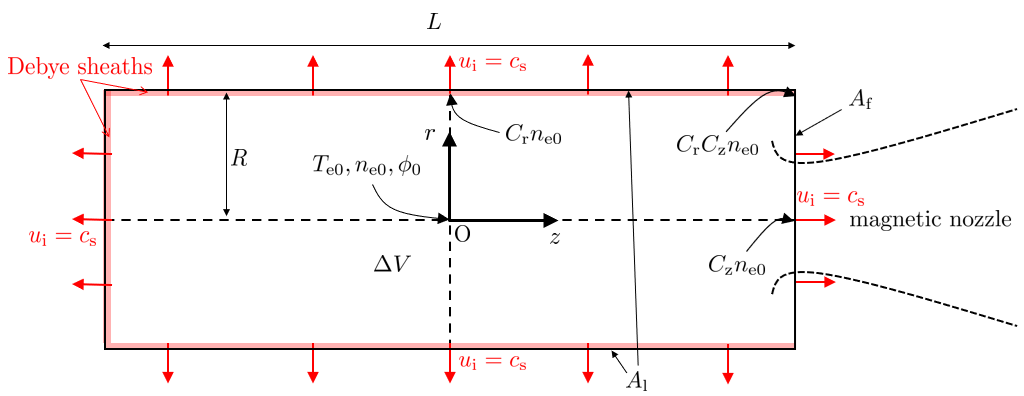}
  \caption{Geometry of the ionization chamber, highlighting the considered plasma parameters and the quasi-neutral plasma density values at the most relevant locations.}
  \label{fig: ionization chamber scheme}
\end{figure}

Referring to Fig.\,\ref{fig: ionization chamber scheme}, and assuming that the plasma is singularly ionized, quasi-neutral everywhere (except at the thin Debye sheaths close to the walls), and that ions exit from the ionization chamber with the Bohm's velocity (or sonic velocity), we can estimate the average plasma density at the exit section $\bar{n}_\mathrm{ef}$ as:

\begin{equation}
\dot{m_\mathrm{i}} = m_\mathrm{i}\bar{n}_\mathrm{ef} c_\mathrm{s} A_\mathrm{f} = m_\mathrm{i} \bar{n}_\mathrm{ef} A_\mathrm{f} \sqrt{\frac{T_\mathrm{e0}}{m_\mathrm{i}}},
\label{eq: mass flow versus density}
\end{equation} 
where $c_\mathrm{s} = \sqrt{\frac{T_\mathrm{e0}}{m_\mathrm{i}}}$ is the ion sonic velocity and $A_\mathrm{f} = \pi R^2$ is the exit surface area. This 0-D model assumes symmetric and quadratic decays along both $z$ and $r$ of the plasma density $n_\mathrm{e}$ from the chamber center. This means that:

\begin{align}
\begin{split}
n_\mathrm{e} (r,z) & = n_\mathrm{e0} \left[ \left(\frac{1}{(1-C_\mathrm{r})}-\frac{r^{2}}{R^{2}}\right)\left( 1- C_\mathrm{r}  \right)\right] \left[\frac{4}{L^{2}}\left( C_\mathrm{z} -1 \right)z^{2} +1 \right]  = \\
& = n_\mathrm{e0} \left[ \frac{4}{L^{2}}\frac{\left(C_\mathrm{z} - 1 \right) \left(C_\mathrm{r} -1 \right)}{R^{2}}r^{2}z^{2} + \frac{\left( C_\mathrm{r} - 1 \right) }{R^{2}}r^{2} +  \frac{4}{L^{2}}\left(C_\mathrm{z} - 1\right)z^{2} + 1\right],
\label{eq: density profiles}
\end{split}
\end{align}
where $n_\mathrm{e0}$ is the plasma density at the ionization chamber center ($r=0,z=0$), and $C_\mathrm{r},C_\mathrm{z}$ are constants that can be obtained from 1D PIC simulations of the plasma-wall interaction. In particular, for Ar ions and a collisionless plasma with no electron emission from the walls, the  ratio between the sheath edge density and the peak density at the symmetry point (cylinder axis) are assumed to be $C_\mathrm{z}=0.5$ (which corresponds to an unmagnetized scenario case) and $C_\mathrm{r}=0.35$. 

The following averaged plasma densities are then introduced: the chamber averaged plasma density $\bar{n}_\mathrm{e}$, the averaged lateral density $\bar{n}_\mathrm{e,l}$ at $r=R$, and the exit section averaged density $\bar{n}_\mathrm{e,f}$ at $z=L/2$, computed as:

\begin{align}
\bar{n}_\mathrm{e} = & \frac{1}{V}\iint_{r,z} n_\mathrm{e}(r,z)2\pi r drdz = 2n_\mathrm{e0}\left[ \frac{\left(  C_\mathrm{z} - 1 \right) \left( C_\mathrm{r} - 1 \right)}{12} + \frac{\left(  C_\mathrm{r} - 1 \right)}{4}  +\frac{\left(  C_\mathrm{z} - 1 \right)}{6}  + \frac{1}{2}  \right],
\label{eq: volume averaged density} \\
\bar{n}_\mathrm{e,l} = & \frac{1}{A_\mathrm{l}}\int_{z=-L/2}^{L/2} n_\mathrm{e}(r=R,z)2\pi R dz = C_\mathrm{r} \left[ \left(  C_\mathrm{z} -1 \right)\frac{1}{3}  + 1\right] n_\mathrm{e0},
\label{eq: sheath edge averaged density} \\
\bar{n}_\mathrm{e,f} = & \frac{1}{A_{f}}\int_{r=0}^R n_\mathrm{e}(r, z=L/2) 2 \pi r dr =  C_\mathrm{z} \left[ \left(  C_\mathrm{r} -1 \right)\frac{1}{2}  + 1\right] n_\mathrm{e0},
\label{eq: exit section density}
\end{align}
where $\Delta V$ is the volume of the ionization chamber, and $A_\mathrm{l} = 2 \pi R L$ is its total lateral surface. Therefore, combining Eqs.\,\ref{eq: mass flow versus density} and \ref{eq: exit section density}, we can obtain the peak plasma density $n_\mathrm{e0}$ as a function of the required mass flow rate:

\begin{equation}
    n_\mathrm{e0} = \frac{1}{ C_\mathrm{z} \left[ \left(  C_\mathrm{r} -1 \right)\frac{1}{2}  + 1\right]} \sqrt{ \frac{m_\mathrm{i}}{T_\mathrm{e0}} } \frac{\dot{m}_\mathrm{i} }   {m_\mathrm{i} A_\mathrm{f}}.
\end{equation}




An average neutral density 
can be obtained by equating the ions production rate in the volume with the rate of ion loss through the lateral and frontal boundary surfaces of the ionization chamber. Such an equilibrium requires that:

\begin{equation}
\bar{n}_\mathrm{e}\bar{n}_\mathrm{n} R_\mathrm{ion}(T_\mathrm{e0})\Delta V = c_\mathrm{s}\left( \bar{n}_\mathrm{e,l} A_\mathrm{l} + 2 \bar{n}_\mathrm{e,f} A_\mathrm{f} \right)
\end{equation}
where $R_\mathrm{ion}=<\sigma_\mathrm{ion}v_\mathrm{e}>$ is the ionization rate, depending on the electron temperature, that can be found by integrating the ionization cross section over a Maxwellian electron distribution. In particular, the ionization cross sections are obtained from Ref.\,\citen{haya03}.

The electron temperature inside the chamber is estimated by solving a power balance equation in which the source power is the RF power deposited in the plasma from helicon waves, and the dissipated power is the power lost from the plasma through the three main dissipation mechanisms: the particle loss on the walls, the ionization of the gas and the excitation of neutrals. The first term is dominant and can be estimated through a fitting law obtained from 1D PIC simulations of the plasma wall interaction in a partially magnetized scenario like this one (where electrons are magnetized but ions are not). In particular, knowing the mean plasma density at the sheath edge $\bar{n}_{e,l}, \bar{n}_{e,f}$ (depending on the considered surface), it is possible to estimate this power as:

\begin{equation} 
P_\mathrm{wall} = 2 A_\mathrm{f} n_\mathrm{e,f} c_\mathrm{s} \gamma_{tot,f} T_{e0} + A_\mathrm{l} n_\mathrm{e,l} c_\mathrm{s} \gamma_{tot,l} T_{e0}   
\end{equation}
where $\gamma_\mathrm{tot}$ represents a total energy transfer coefficient (summing the contribution of both electrons and ions) that depends on the considered surface and more precisely on the angle $\psi$ of the magnetic induction field relative to the surface normal direction. In particular, $\psi=0$ for the two frontal surfaces, and $\approx 89^o$ for the lateral surfaces (nearly parallel incidence). In general, for cold ions ($T_\mathrm{i0}=0$), the following fitting can be used:

\begin{equation}  
\gamma_\mathrm{tot} = \frac{5}{2} - \ln \left(\dfrac{\sqrt{\frac{2\pi m_\mathrm{e}}{m_\mathrm{i}}} C_\gamma(\psi,r_\mathrm{Le}/\lambda_\mathrm{De}, m_\mathrm{e}/m_\mathrm{i})}{\cos\psi}\right)
\end{equation}
where $C_\gamma$ is a parameter that, in general, depends on the magnetic field angle, on the ratio between the electron Larmor radius and the Debye length $\lambda_\mathrm{De}$, and on the electron to ion Larmor radius. In a plasma with Ar ions and an electron Larmor radius comparable with the Debye length ($\lambda_\mathrm{De}/r_\mathrm{Le}=O(1)$), we obtain $C_\mathrm{\gamma,f}=1$ for the frontal surfaces ($\psi=0$), and $C_\mathrm{\gamma,l}=0.4$ for the lateral surface ($\psi \approx 89^o$). Again, the above fittings and coefficients have been obtained from 1D PIC simulations of a non-emitting wall surface interacting with a collisionless plasma, so that the effects of secondary electrons and charged particles collisions with neutrals are neglected.

Concerning the other two loss mechanisms, the expressions of the ionization and excitation power are the following:

\begin{equation} 
P_\mathrm{exc} = \bar{n}_{n}\bar{n}_{e} R_\mathrm{exc}(T_\mathrm{e0})E_\mathrm{exc} \Delta V,
\end{equation}

\begin{equation} 
P_\mathrm{ion} = \bar{n}_{n}\bar{n}_{e} R_\mathrm{ion}(T_\mathrm{e0})E_\mathrm{ion} \Delta V,
\end{equation}
where the rates are obtained from integrating over a Maxwellian electron distribution the cross sections available in Refs.\citen{haya03,lxcat} for respectively ionization and excitation.
The ionization and the excitation energy thresholds are respectively 15.76 and 11.5 eV for argon. 

The equilibrium temperature $ T_{e0} $ of the helicon thruster is then computed by assuming a given absorbed power and equating it to the total power losses. This means that:

\begin{equation}
P_\mathrm{\Omega} = P_\mathrm{wall} (T_\mathrm{e0})+ P_\mathrm{ion} (T_\mathrm{e0}) + P_\mathrm{exc} (T_\mathrm{e0}),
\end{equation}
where the terms on the right hand side depend on the electron temperature.
The targeted absorbed power is then the requested input for the finite-element full-wave model (described in the following section), which has the goal of validating the design of a helicon antenna capable of delivering that power to the plasma. 

The helicon antenna is designed according to the helicon wave theory \cite{chen91}. The helicon antenna is a Nagoya type-III antenna which accelerates free electrons in a neutral gas immersed in a magnetostatic field, ionizing it through energy deposition. 
It is assumed, as stated in the previous equation, that the deposited energy in the plasma by helicon waves is a fraction of the RF input power and heat the plasma through Joule heating. The size of the helicon antenna is determined by combining the quadrature relation for the total wave number $k$ of the helicon wave and the plasma dispersion relation for helicons in a magnetized plasma, as follows:
\begin{equation}
\begin{cases}
  k^{2} = k_{\perp}^{2} + k_{\parallel}^{2}, \\
   k_{\parallel} = \dfrac{\omega}{k}\dfrac{\omega_\mathrm{p,e}^{2}}{\omega_\mathrm{c,e}c^{2}},
\end{cases}
\end{equation}
where $\omega_{c,e}$ is the electron cyclotron frequency, $c$ is the speed of light and $\omega_{p,e}$ is the electron plasma frequency. From this linear system, the wave numbers of the helicon wave are computed, such as the total wave number $k$, the parallel wave number $k_{\parallel}$  and the transversal wave number $k_{\perp}$. 
Then, the size of the antenna is determined by assuming that the antenna is resonant with an helicon mode propagating in the magnetized plasma. The antenna is assumed of cylindrical shape and therefore it is determined from the antenna radius $a$ and the antenna length $L_\mathrm{a}$. The antenna radius $a$, is computed from the ionization chamber radius $R$, assuming a thickness of the ionization chamber equal to $5 \hspace{1 mm}$ mm. The design conditions for the Nagoya type-III antenna are 

\begin{equation}
\begin{cases}
  k_{\perp} = \dfrac{3.83}{R}, \\
  L_\mathrm{a} = \dfrac{\lambda}{2}.   \\
\end{cases}
\end{equation}

Therefore, once the antenna shape $(a,L_\mathrm{a})$ is chosen and the chamber averaged plasma density $\bar{n}_\mathrm{e}$ (Eq.\,\ref{eq: volume averaged density}) is known, the required magnetostatic field can be computed from the plasma dispersion relation according to the following equation:

\begin{equation}
B_{0} = \dfrac{\omega}{k_{\parallel}} \dfrac{\mu_{0} e \bar{n}_\mathrm{e}}{k}  
\label{eq: magnetostatic field}
\end{equation}

and therefore the main output parameters of the 0D model 
are fully determined.


\section{Full-wave simulation model for the ionization chamber}
\label{sec: plasma wave model}

The plasma in the ionization chamber is simulated with a in-house 3D FEM code which solves Maxwell's equations for a realistic 3D geometry in a cartesian reference frame. The obtained full-wave solutions allow to compute the electromagnetic field coupled by the helicon antenna in the plasma and to estimate the power transferred from the helicon antenna to the plasma. The simulation model assumes that a feeding power (as requested from the 0D model) is provided to the helicon antenna and a fraction of this power is absorbed by the plasma through collisional damping. 

The plasma parameters (plasma density, the neutrals density, the electron temperature) and the geometrical parameters (the helicon antenna dimensions and the size of the ionization chamber) are selected in accordance with the 0D model of the thruster described above. The magnetostatic field, necessary for the helicon wave propagation, is derived from the dispersion relation of helicon waves \cite{chen91} (see Eq.\,\ref{eq: magnetostatic field}).

The plasma, already ionized, is modeled as an equivalent homogeneous anisotropic dielectric with permettivity given by the Stix dielectric tensor \cite{STIX92} with collisional corrections. Electron-ion and electron-neutral collisions are computed according to the following formulas:

\begin{align}
    \nu_\mathrm{ei} = & 3.9 \cdot \hspace{1 mm} 10^{-6}n_\mathrm{i}Z_\mathrm{i}^{2}\frac{\ln\Lambda}{T_\mathrm{e0}^{\frac{3}{2}}}, \\
    \nu_\mathrm{en} = & \sigma_\mathrm{n} n_\mathrm{n} v_\mathrm{th,e},
\end{align}
where $Z_\mathrm{i}=1$ is the ion charge number, $\ln{\Lambda}$ is the Coulomb logarithm, $\sigma_\mathrm{n}$ is a constant neutral atom cross section, and $v_\mathrm{th,e}=\sqrt{T_{e0}/m_{e}}$ is the electron thermal velocity in the ionization chamber. For the considered case, $\nu_\mathrm{ei}\approx 3.9$ MHz, and $\nu_\mathrm{en}\approx 4.2$ MHz. Such frequencies are considered in the plasma dielectric tensor using correction terms proportional to the collisional frequency \cite{card14,mela15}. The helicon electric field and the induced current density in the plasma are obtained from the full-wave simulation results and they allow the computation of the deposited power in the plasma:
\begin{equation}
 P_\Omega (T_{e}) = \frac{1}{2}\int_{V}\mathrm{Re}\left(\bm{j} \cdot \bm{E}^{*}\right) \mathrm{d}V
\end{equation}
where $\bm{j}$ and $\bm{E}^{*}$ are complex vectors representing the current density vector and the complex conjugate of the helicon electric field vector and $V$ is the volume of the ionization chamber. The deposited power $ P_{\Omega} $ is the source power in the power balance equation used in the 0-D model for the electron temperature evaluation (Eq. 16).

The main results obtained from the full-wave simulation are the verification of the helicon antenna design for the target absorbed power (as required by the 0D model) and the computation of the exact RF input power level required for the desired operative point of the antenna. Last, but not least, the verification of the RF coupling with a propagating helicon mode in the target magnetized plasma.

\section{Particle-in-cell model of the magnetic nozzle expansion}
\label{sec: PIC model}
The collisionless expansion within the magnetic nozzle is simulated with an electrostatic particle-in-cell code \cite{tacc23} named PICCOLO \cite{tacc22}. This choice is justified by the fact that most of the electromagnetic wave absorption takes place inside the ionization chamber since the helicon wave becomes evanescent in the nozzle, not far away from the exit surface of the chamber.
Ion and electron macro-particles (each one representing a large number of elementary particles, here referred to as macro-particle weight) are moved according to Newton's equation and the local electric and magnetic fields $\bm{E}$ and $\bm{B}$:
\begin{equation}
    m \frac{d\bm{v}}{\mathrm{d}t} = q \left( \bm{E} + \bm{v} \times \bm{B} \right),
    \label{eq: newton}
\end{equation}
where $m,q$ are the elementary particle mass and charge. A classical leap-frog algorithm \cite{BIRD91} is used to advance particles in time. 

While the magnetic induction field is given as an input map (from the 0D global model and the antenna design constraints) and is fixed throughout the simulation, the electrostatic field $\bm{E}= - \nabla \phi$ is obtained by differentiation of the self-consistent electric potential $\phi$, which is the solution of Poisson's equation in the considered cylindrical coordinates $r$ and $z$:
\begin{equation}
    \frac{1}{r}\frac{\partial  \phi}{\partial r} + \frac{\partial^2 \phi}{\partial r^2} + \frac{\partial^2 \phi}{\partial z^2} = - \frac{e}{\epsilon_0} \left( n_\mathrm{i}-n_\mathrm{e} \right),
\end{equation}
where $n_\mathrm{i}$ represents the singly charged Ar ion density, and $\epsilon_0$ is the vacuum permittivity. 
Neither neutral particles/background nor charged particle collisions are included in the simulations shown here (collisionless plasma).
\begin{figure}[ht!]
  \centering
  \includegraphics[width = 0.9\textwidth]{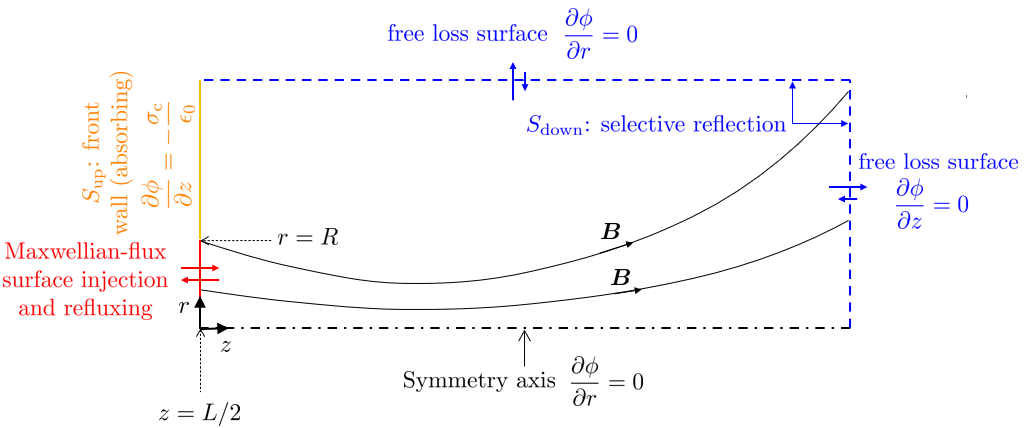}
  \caption{Particle-in-cell simulation scheme for the magnetic nozzle, indicating the boundary conditions for both particles and electric potential. The injection region extends from $r=0$ to $r=R$. Note that the depicted magnetic field is only qualitative, and, simulations only consider the region where it is fully divergent.}
  \label{fig: MN sim scheme}
\end{figure}
The boundary conditions for ion and electron macro-particles are then shown in Fig.\,\ref{fig: MN sim scheme}. 
In particular, the domain starts at the exit section of the ionization chamber (with a radius $R$), and an ambipolar flux of ions and electrons is injected from a region extending from $r=0$ to $r=R$ with a Maxwellian-flux velocity distribution\cite{cich23a}, and a spatial distribution given by the parabolic radial profile of Eq.\,\ref{eq: density profiles}. At $z=L/2$ and for $r>R$, a dielectric material surface is assumed, meaning that ions and electrons are lost there and accounted for in the local surface charge density $\sigma_\mathrm{c}$. While quasi-neutrality is automatically satisfied at the injection surface by refluxing all particles that cross the injection surface ($z=L/2$, $r<R$) towards the source, at the open free-loss boundaries (dashed blue lines) all ions are lost and electrons are either reflected or not, according to their mechanical energy $E_\mathrm{tot}$ (relative to infinity),
\begin{equation}
    E_\mathrm{tot} = \frac{1}{2} m_\mathrm{e} v_\mathrm{e}^2 - e \left( \phi_\mathrm{exit} - \phi_\mathrm{\infty} \right),
\end{equation}
where $v_\mathrm{e}$ is the electron velocity, $\phi_\mathrm{exit}$ is the local electric potential at the outflow position, and $\phi_\mathrm{\infty}$ is the potential at infinity.
All electrons with $E_\mathrm{tot}<0$ are specularly reflected backwards toward the simulation domain, while all electrons with $E_\mathrm{tot}>0$ are lost and charge up a virtual capacitor at infinity, according to:
\begin{equation}
 \frac{\mathrm{d} \phi_\infty}{\mathrm{d}t} = \frac{ I_\mathrm{i\infty} + I_\mathrm{e\infty}}{C_\infty},
 \label{eq: infinity potential}
\end{equation}
where $I_\mathrm{i\infty},I_\mathrm{e\infty}$ are the ion and electron currents that leave the domain to infinity, and $C_\infty$ is an equivalent electric capacity.

\section{Simulation results}
\label{sec: simulation results}

\subsection{0-D model sizing results}

The global design model assumes a target propulsive performance of the helicon plasma thruster, with an achieved thrust (12 mN) and specific impulse (1200 s), and with an absorbed plasma power of around 1 kW. The main parameters of the 0D global sizing model are reported in Tab. \,\ref{tab: global model parameters}. The reported data are split in inputs and outputs of the model. 
The geometry of the antenna and ionization chamber are chosen to obtain the above mentioned target absorbed power. 
The results of the 0-D model are the required inputs for both the full-wave and PIC simulations. In fact, the thermodynamic state in terms of pressure, density and temperature of both plasma and neutrals, and moreover the size of the helicon plasma thruster and the magnetostatic field are provided as input to the electromagnetic simulation software to define the geometry of the thruster and to compute the helicon mode solution and the power transfer between the helicon antenna and the plasma. In particular, a magnetostatic field of around 90 G is required and in this case it is generated by two coils of radius 12 cm located axially at the two ends of the ionization chamber ($z=\pm L/2$).
The efficiency of the designed thruster is quite low (3\%), although it is here emphasized that the 0-D model provides a conservative estimate of this figure, and higher achievable values may be obtained.
The thruster efficiency is computed as: 
\begin{equation}
    \eta_\mathrm{t} = \frac{F g_0 I_\mathrm{sp}}{2P_\mathrm{RF}}
\end{equation}
where $P_\mathrm{RF}$ is the total RF power provided in input to the helicon antenna, which relates to the absorbed plasma power according to $P_\Omega = \eta_\mathrm{RF} P_\mathrm{RF}$ where $\eta_\mathrm{RF}$ is the RF power transmission efficiency (from RF amplifier to plasma). Here a conservative estimate equal to 70\% is assumed. Note that $\eta_\mathrm{t}$ is the total thruster efficiency, considering the input power at the RF amplifier while, in some cases, the thrust efficiency is only referred to the thruster chamber (with the absorbed plasma power in the denominator).

\begin{table}[!ht]
    \centering
    \begin{tabular}{|c|c|c|c}
     \hline
     \hline
     Input parameters & units & values \\
     \hline
     \hline
     Chemical species & (-) & Argon \\
     Feeding RF frequency &  MHz  & 13.56 \ \\
     Thrust, $F$ & mN  & 12\\
     Specific impulse, $I_\mathrm{sp}$ & s & 1200 \\
     Propellant utilization efficiency, $\eta_{m}$ & (-) & 0.85 \\
     RF power transmission efficiency, $\eta_\mathrm{RF}$ & (-) & 0.70 \\
     Ionization chamber radius, $R$ & cm  & 3  \\
     Helicon antenna radius, $a$ & cm & 3.5 \\
     Helicon antenna length, $L_\mathrm{a}$ & cm  & 12 \\
     Ionization chamber length, $L$  & cm & 12   \\
     Neutrals temperature & K & 300 \\
     \hline
     \hline
     Output parameters & units & values \\
     \hline
     \hline
     Electron temperature, $T_\mathrm{e0} $ & eV & 7.667 \\
     Ion mass flow rate, $\dot{m}_\mathrm{i}$ & mg/s & 0.866 \\
     Peak electron density, $n_\mathrm{e0}$& m$^{-3}$ &  $3.181 \cdot 10^{18}$ \\
     Mean electron density, $\bar{n}_\mathrm{e}$ & m$^{-3}$ & $1.790 \cdot 10^{18}$\\
     Mean neutral density, $\bar{n}_\mathrm{n}$ & m$^{-3}$& $2.256 \cdot 10^{19}$ \\
     Magnetostatic field, $B_{0}$ & G & 90 \\
     Deposited power in the plasma, $P_\Omega$ & W & 1147.2\\
     Total thrust efficiency, $\eta_\mathrm{t}$ & (-) & 4.3 \% \\
     \hline
     \hline
    \end{tabular}
    \caption{Parameters considered in the global model for sizing the helicon thruster.}
    \label{tab: global model parameters}
\end{table}

\subsection{Full-wave simulation of the plasma-wave interaction}

The full-wave simulations are made with an in-house electromagnetic 3D FEM code that, for a given geometry, is capable of simulating the electromagnetic wave field excited in the plasma by the helicon antenna. The geometry considered here is the Nagoya type-III antenna that is fed by a coaxial cable connecting the RF generator with the antenna feeding port, a spatial gap in one of the two antenna legs. When fed by a RF source, the Nagoya type-III antenna can excite an electromagnetic plasma mode corresponding to a propagating helicon wave with poloidal number $m = 1$ \cite{chen91}. While the helicon wave is propagating, the wave deposits its power in the plasma through collisional damping and the plasma temperature is determined by Joule heating. As stated previously, the heating power is computed from the volume integral of the scalar product between the current density vector and the complex conjugate of the electric field vector. 

\begin{figure}[ht!]
  \centering
  \includegraphics[width = 0.6\textwidth]{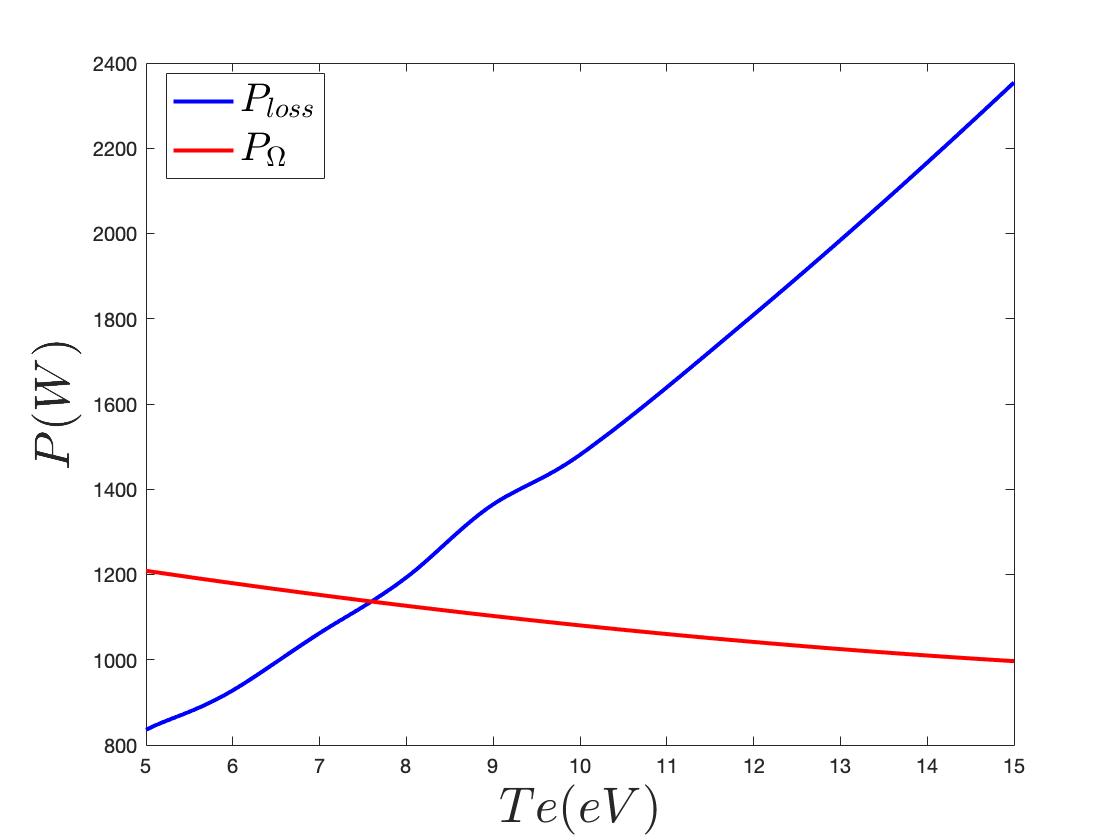}
  \caption{Graphical solution of the plasma power balance equation for the helicon thruster with an input RF power of about 2 kW}
  \label{fig: power balance}
\end{figure}

The operational point of the thruster is obtained fine tuning the RF input power at about 2 kW at the target electron temperature and it corresponds to the intersection between the deposited RF power and the total loss power when they are computed as a function of the electron temperature, as shown in Figure \ref{fig: power balance}. Indeed, the intersection of the two curves is the steady state operational point of the helicon plasma thruster at the target electron temperature (as computed from the 0D model), i.e. where the deposited power in the plasma 
is equal to the total power lost.

\begin{figure}[ht!]
  \centering
  \includegraphics[width = 0.6\textwidth]{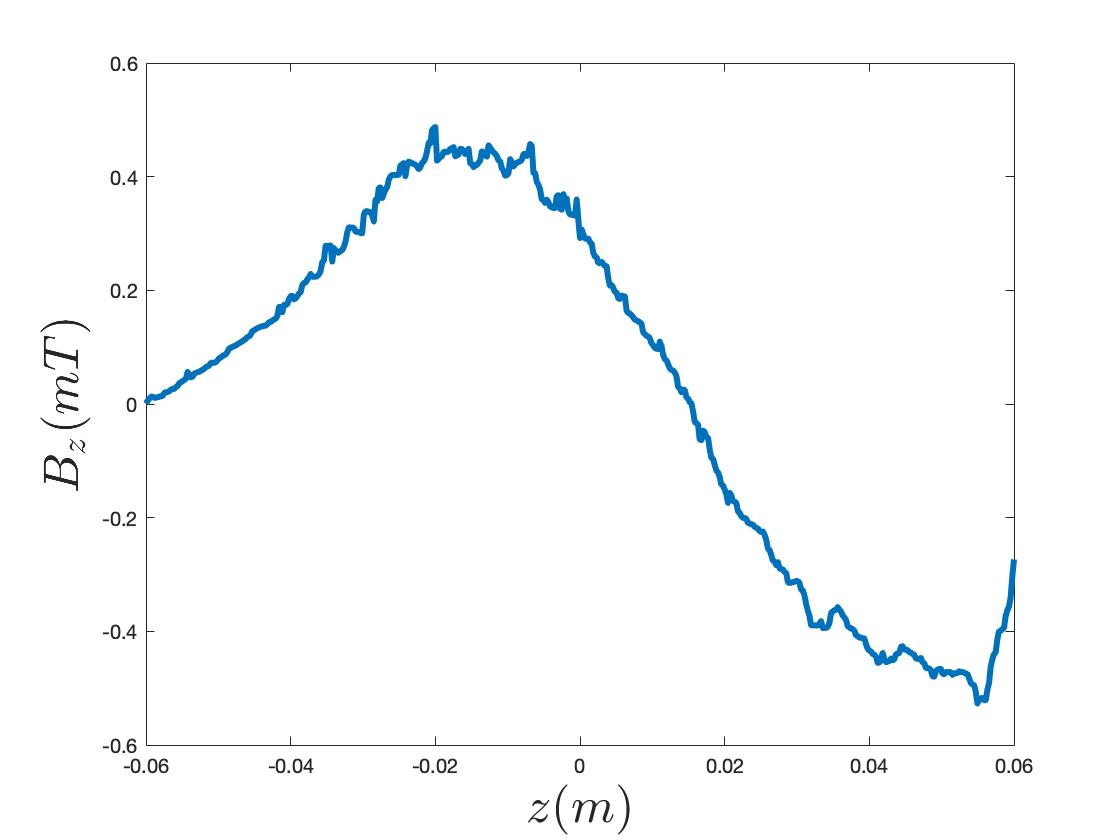}
  \caption{Longitudinal oscillation of the axial component of the helicon magnetic field (Bz)}
  \label{fig: magnetic field evolution}
\end{figure}

\subsection{Full particle-in-cell simulation of the magnetic nozzle}

The PIC simulation of the magnetic nozzle considers the parameters summarized in Tab.\,\ref{tab: PIC parameters}. 
\begin{table}[!ht]
    \centering
    \begin{tabular}{|c|c|c|c}
     \hline
     \hline
     Parameter & units & values \\
     \hline
     \hline
     Injected ion mass flow  & mg/s & 0.866 \\
     Dielectric constant scaling factor   &  (-) & 9 \\
     PIC time step           &  s  & 5 $\cdot 10^{-12}$  \\
     Elementary electron mass & kg  & 9.109 $\cdot 10^{-31}$ \\
     Elementary ion mass & kg  & 6.633 $\cdot 10^{-26}$ \\
     Macro-particle weight    & (-) & 4.68 $\cdot 10^6$  \\
     Virtual infinity capacity $C_\mathrm{\infty}$           &  nF  & 0.25  \\
     Radial extension of domain & cm & 15 \\
     Axial extension of domain & cm & 25 \\
     Cell size along $z$ and $r$ & $\mu$m & 50 \\ Injection ion axial fluid velocity & m/s & 4303 \\
     Injection electron temperature & eV & 7.667 \\
     Injection ion temperature & eV & 0.5 \\
     \hline
     \hline
    \end{tabular}
    \caption{Parameters considered for the PIC simulation of the magnetic nozzle expansion.}
    \label{tab: PIC parameters}
\end{table}
The time step and the cell size are chosen to respect the PIC constraints on Debye length, plasma frequency and cyclotron frequency \cite{tacc23}. Ions are injected with sonic conditions with a fully axial fluid velocity, and a temperature of 0.5 eV, resembling a realistic velocity dispersion due to the fact that they are created at different positions inside the ionization chamber. An artificially increased permittivity by a factor of 9 is used to reduce the simulation cost, which corresponds to simulating a Debye length 3 times as large as the real one. This, however, should not impact significantly the obtained results, since the plasma is quasi-neutral nearly everywhere and the total magnetic nozzle potential drop is not affected by the value of $\epsilon_0$, as shown in Eq.\,\ref{eq: delta phi drop}.

\begin{figure}[ht!]
  \centering
  \includegraphics[width = 0.6\textwidth]{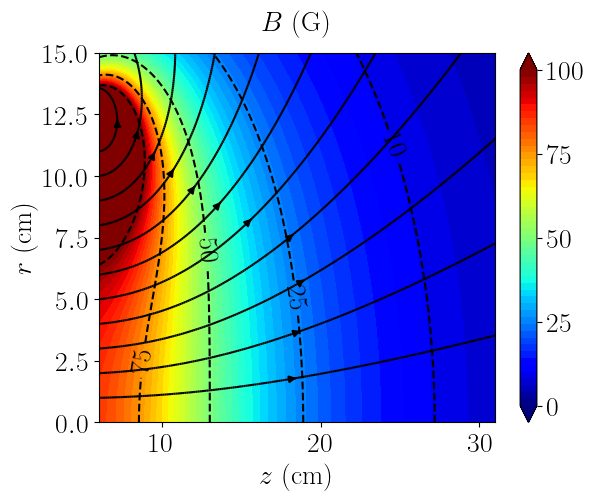}
  \caption{Magnetic induction field intensity and streamlines.}
  \label{fig: B field map PIC}
\end{figure}

The magnetic field intensity and streamlines is shown in Fig.\,\ref{fig: B field map PIC}. 
At the exit surface, the intensity is slightly lower than 90 G, and decreases to just a few G downstream. The intensity peaks at the exit surface at $r=12$ cm, as that is the radius of one of the two coils that generate this magneto-static field (the other coil is located at $z=-L/2$, i.e. at the other end of the ionization chamber).

\begin{figure}[ht!]
  \centering
  \includegraphics[width = 0.6\textwidth]{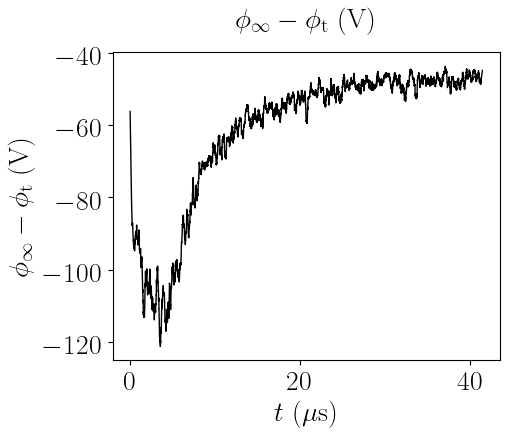}
  \caption{Infinity potential evolution relative to the nozzle throat (located at $z=6$ cm, $r=0$). This potential difference is time-averaged over a window of 0.25 $\mu$s.}
  \label{fig: infinity potential evo}
\end{figure}

The infinity potential evolution with time is shown in Fig.\,\ref{fig: infinity potential evo}. 
In particular, this represents the infinity potential (Eq.\,\ref{eq: infinity potential}) relative to the nozzle throat ($r=0, z = L/2$) and its oscillations are mainly due to the PIC noise at nozzle throat (the centerline in axisymmetric simulations is always affected by a large PIC noise, see Ref.\,\citen{domi18}). Nevertheless, the absolute drop from nozzle throat to infinity should tend asymptotically to the one predicted by Eq.\,\ref{eq: delta phi drop}, which is around 40 V in this case. In this case, the predicted drop is slightly larger than expected (approx. 50 V), since Eq.\,\ref{eq: delta phi drop} underestimates the total potential drop across the magnetic nozzle, as also confirmed in other studies \cite{ahed20}.

\begin{figure}[ht!]
  \centering
  \includegraphics[width = 1\textwidth]{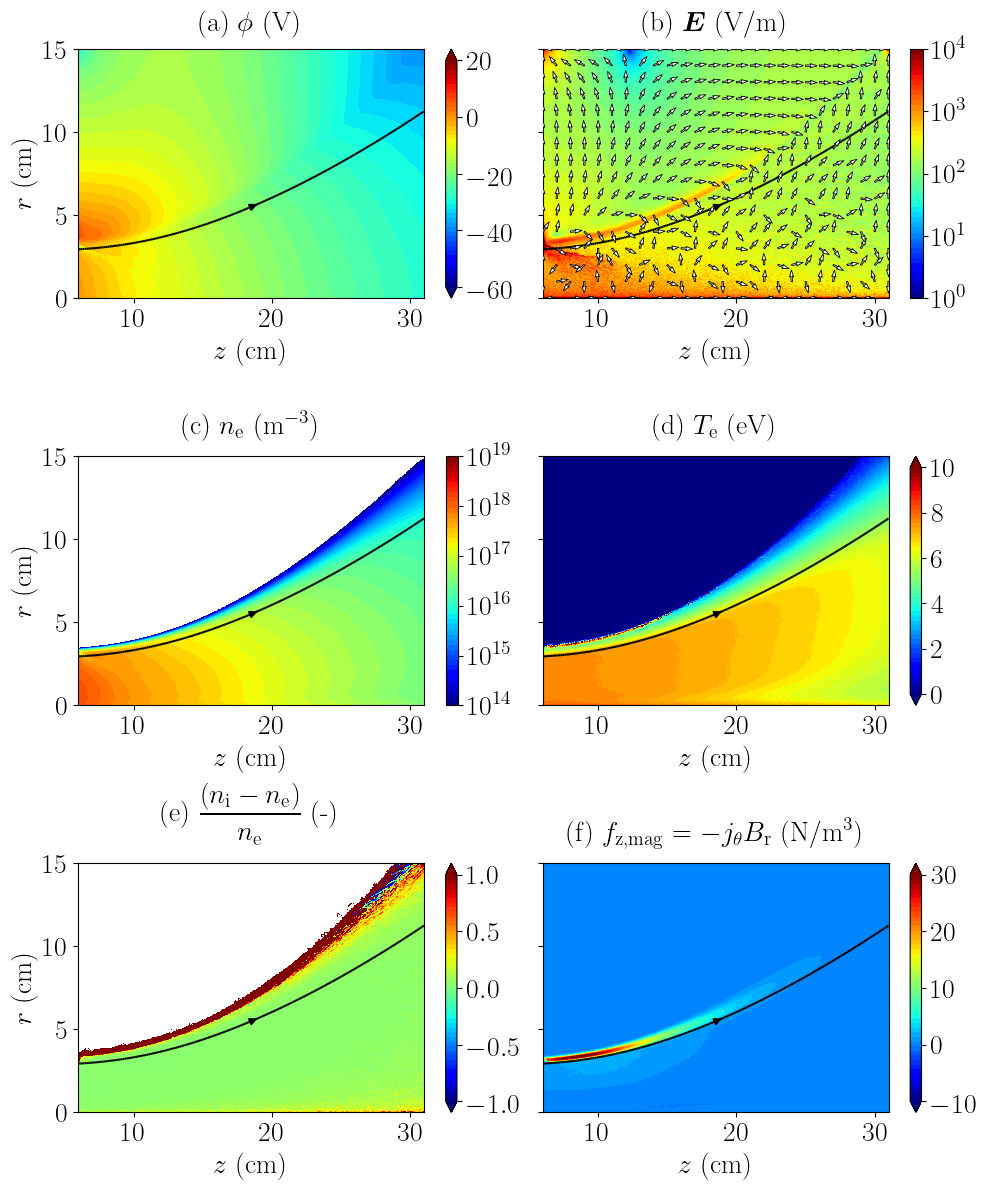}
  \caption{2D plasma properties maps, showing (a) electric potential, (b) electric field magnitude with direction, (c) electron density, (d) electron temperature, (e) relative charge density $\left(n_\mathrm{i}-n_\mathrm{e}\right)/n_\mathrm{e}$, and (f) axial magnetic force density. The electric potential is set to 0 at the nozzle throat ($z=L/2,r=0$). The outermost magnetic field line from the ionization chamber (through $z=L/2,r=R$) is shown by a black line. Results are time-averaged over a window of 0.25 $\mu$s and refer to a simulation time of 41.5 $\mu$s.}
  \label{fig: MN 2D plasma maps}
\end{figure}

The 2D maps of several variables of interest are finally reported in Fig.\,\ref{fig: MN 2D plasma maps}. 
The electrostatic potential and field of subplots (a,b) clearly show the effect of the magnetic nozzle, which is that of reducing the divergence of the emitted plume by focusing the ions axially. A plume boundary effect can be observed in both subplots and tends to propagate downstream as well. This effect is produced by both the injection strategy (purely axial ion injection velocity) and by the abrupt density drop from $C_\mathrm{z}C_\mathrm{r}n_\mathrm{e0}$ to 0 at $r=R, z= L/2$.

Fig.\,\ref{fig: MN 2D plasma maps} (c,d) show the electron density and temperature in the meridional plane. Clearly, since electrons are magnetized and the plasma is collisionless, they cannot traverse radially the magnetic field, so that their density drops to 0 beyond the most external magnetic field line (from the inner ionization chamber), shown by a black solid line. The electron temperature, on the other hand, presents a minimum close to the plume boundary (similarly to what happens in a classical Debye sheath) and slowly decreases within the magnetic nozzle expansion, consistently with the expectations that electron thermal energy is converted into ion directed kinetic energy.

Fig.\,\ref{fig: MN 2D plasma maps} (e) then shows the relative charge density, highlighting the non-neutral regions of the expansion. A positively charged region can be observed close to the plume boundary due to the fact that ions are essentially unmagnetized and can easily traverse magnetic field lines, while electrons cannot. In the near totality of the core plasma, however, quasi-neutrality holds well, including at the injection boundary $z=L/2$, thus validating the simulated injection approach.

Finally, Fig.\,\ref{fig: MN 2D plasma maps} (f) shows the axial magnetic thrust force density $\left(\bm{j} \times \bm{B}\right)_\mathrm{z}$, i.e. the magnetic thrust per unit volume. Since the magnetic induction field has no azimuthal component, this is given by $f_\mathrm{z,mag} = - j_\theta B_\mathrm{r}$, and is mainly generated at the peripheral plume regions. Peak values close to 30 N/m$^3$ can be observed. Since the plasma plume is essentially quasi-neutral everywhere (see Fig.\,\ref{fig: MN 2D plasma maps}(e)), this volumetric force is the only direct contribution to thrust of the magnetic nozzle.

The total thrust force is obtained by integrating the total axial momentum flow through the simulation boundaries (with the exception of the injection boundary $r<R, z=L/2$) as:

\begin{align}
\begin{cases}
    F = & \iint_{S_\mathrm{down}} {\left[ \left( n_\mathrm{i} \bm{u}_\mathrm{i} \cdot \bm{1}_\perp \right) m_\mathrm{i} u_\mathrm{z,i} 
    + \left( n_\mathrm{e} \bm{u}_\mathrm{e} \cdot \bm{1}_\perp \right) m_\mathrm{e} u_\mathrm{z,e} + \left(p_\mathrm{i}+p_\mathrm{e} \right) \bm{1}_\perp \cdot \bm{1}_\mathrm{z} \right] \mathrm{d}S} \\
    & + \iint_{S_\mathrm{up}} \left[ \left( n_\mathrm{i} 
    \bm{u}_\mathrm{i} \cdot \bm{1}_\perp \right) m_\mathrm{i} u_\mathrm{z,i} 
    + \left( n_\mathrm{e} 
    \bm{u}_\mathrm{e} \cdot \bm{1}_\perp \right) m_\mathrm{e} u_\mathrm{z,e} + \left(p_\mathrm{i}+p_\mathrm{e} \right) \bm{1}_\perp \cdot \bm{1}_\mathrm{z}  \right] \mathrm{d}S,
    \end{cases}
\end{align}
where $\bm{1}_\perp$ is the unit vector normal to the boundaries and directed outward (away from the simulation domain), $\bm{1}_\mathrm{z}$ is the unit vector along the axial direction $z$, $S_\mathrm{down}$ is the downstream open boundary surface, and $S_\mathrm{up}$ is the upstream back-flow surface ($r>R, z=L/2$), as shown in Fig.\,\ref{fig: MN sim scheme}. Note that for the second integral, both the average axial momentum and pressure contributions are negative, since $\bm{1}_\perp = - \bm{1}_\mathrm{z}$. After a simulation time of 41.5 $\mu$s, the total thrust force amounts to 10.9 mN and is almost completely stationary. This can be split between 7.6 mN of injected momentum flow (from the ionization chamber), 3.4 mN due to the integrated axial magnetic force density over the simulated volume ($\iiint_\mathrm{vol} -j_\theta B_\mathrm{r} \mathrm{d} V$), and a small negative contribution of -0.1 mN due to the integrated electric force ($\iiint_\mathrm{vol} \rho_\mathrm{c} E_\mathrm{z} \mathrm{d} V$).
This thrust value is close to the one predicted by the 0-D global model (12 mN), which did not assume plume divergence losses. These can be estimated as the ratio between the $z$-directed and the total (in any direction) energy flows through the surfaces $S_\mathrm{down}$ and $S_\mathrm{up}$. Neglecting pressure and heat flux effects at the downstream section (where the plasma density has decreased significantly), this plume divergence efficiency can be roughly estimated from the ion fluid properties as:
\begin{equation}
    \eta_\mathrm{div} \approx \dfrac{ \iint_{S_\mathrm{down}} {\left[ 
    n_\mathrm{i} \bm{u}_\mathrm{i} \cdot \bm{1}_\perp m_\mathrm{i} u_\mathrm{z,i}^2 \right] \mathrm{d}S }  - \iint_{S_\mathrm{up}} 
    { \left[ n_\mathrm{i} \bm{u}_\mathrm{i} \cdot \bm{1}_\perp m_\mathrm{i} u_\mathrm{z,i}^2 \right] \mathrm{d}S }  } 
    {\iint_{S_\mathrm{down}+S_\mathrm{up}} {\left[ 
    n_\mathrm{i} \bm{u}_\mathrm{i} \cdot \bm{1}_\perp m_\mathrm{i} u_\mathrm{i}^2 \right] \mathrm{d}S } },
\end{equation}
i.e. as the ratio between the $z$-directed ion kinetic energy flow (with sign) and the total ion kinetic energy flow through the loss surfaces. At the latest available time step, this divergence efficiency is almost stationary and equal to approx. 87\%. Clearly, such a high value can be ascribed to both the magnetic topology and the non-modeled ion/electron collisions with the neutral background, which should act to reduce it slightly.

\section{Conclusions}
\label{sec: conclusions}

In this work, a 0D global model has been developed to size the main components of a helicon thruster and the main operating plasma and magnetic circuit parameters. The sizing model is based on fundamental propulsive and physical models that allow to dimension the helicon thruster for a target performance in terms of thrust and specific impulse. This 0-D model has been coupled with advenced simulations that predict both the helicon mode coupling by the antenna and plasma fields generated inside the thruster ionization chamber and the plasma expansion of the magnetic nozzle, in order to estimate the effective performance of the thruster.

In particular, the electromagnetic simulations of the plasma-wave interaction allow to verify that a propagating helicon mode is coupled by the antenna to the plasma and to compute the RF input power and the realistic dimensions of the helicon antenna required to meet the desired operative point of the thruster. The electrostatic PIC simulations of the magnetic nozzle expansion, on the other hand, cast light on the acceleration process and permit to estimate the propulsive performance of the thruster. Simulations are in good agreement with the prediction of the 0D global model, both for the electromagnetic and electrostatic parts. Therefore, it can be stated that the proposed 0D model can provide a reasonable preliminary sizing of the helicon thruster.

Based on these results, an optimization study of the helicon thruster geometry and operational conditions will be pursued in a successive work. Additional magnetic nozzle PIC simulations (featuring also particle collisions, neglected here) will permit to optimize the magnetic circuit in terms of the achieved thrust (i.e. the one maximizing the divergence efficiency of the plume). 
At a second level, when a real helicon prototype will be available for vacuum testing, by combining the numerical predictions with real experimental measurements, the 0D global model parameters will be further tuned and, for given constraints, a helicon thruster with an optimal thrust efficiency will be designed.

\bibliographystyle{aiaa}
\bibliography{papers}

\begin{thebibliography}{10}
\newcommand{\enquote}[1]{``#1''}

\bibitem{chen91}
Chen, F.~F., \enquote{Plasma ionization by helicon waves,} {\em Plasma Physics and Controlled Fusion\/}, Vol.~33, No.~4, 1991, pp.~339.

\bibitem{STIX92}
Stix, T.~H., {\em Waves in plasmas\/}, Springer Science \& Business Media, 1992.

\bibitem{card14}
Cardinali, A., Melazzi, D., Manente, M., and Pavarin, D., \enquote{Ray-tracing WKB analysis of Whistler waves in non-uniform magnetic fields applied to space thrusters,} {\em Plasma Sources Science and Technology\/}, Vol.~23, No.~1, 2014, pp.~015013.

\bibitem{mela15}
Melazzi, D. and Lancellotti, V., \enquote{A comparative study of radiofrequency antennas for Helicon plasma sources,} {\em Plasma Sources Science and Technology\/}, Vol.~24, No.~2, 2015, pp.~025024.

\bibitem{tacc23}
Taccogna, F., Cichocki, F., Eremin, D., Fubiani, G., and Garrigues, L., \enquote{Plasma propulsion modeling with particle-based algorithms,} {\em Journal of Applied Physics\/}, Vol.~134, No.~15, 2023.

\bibitem{ahed20}
Ahedo, E., Correyero, S., Navarro-Cavall{\'e}, J., and Merino, M., \enquote{Macroscopic and parametric study of a kinetic plasma expansion in a paraxial magnetic nozzle,} {\em Plasma Sources Science and Technology\/}, Vol.~29, No.~4, 2020, pp.~045017.

\bibitem{haya03}
Hayashi, M., \enquote{Bibliography of electron and photon cross sections with atoms and molecules published in the 20th century,} Tech. rep., National Institute for Fusion Science of Japan, 2003.

\bibitem{lxcat}
Phelps, A.~V., \enquote{A compilation of atomic and molecular data,} https://fr.lxcat.net/data/set\_databases.php.

\bibitem{tacc22}
Taccogna, F., Cichocki, F., and Minelli, P., \enquote{Towards full three-dimensional modeling of Hall thruster internal channel discharges,} {\em 37th International Electric Propulsion Conference\/}, No. IEPC-2022-340, Electric Rocket Propulsion Society, Boston, June 19-23, 2022.

\bibitem{BIRD91}
Birdsall, C.~K. and Langdon, A.~B., {\em Plasma physics via computer simulation\/}, CRC press, 1991.

\bibitem{cich23a}
Cichocki, F., Sciortino, V., Giordano, F., Minelli, P., and Taccogna, F., \enquote{Two-dimensional collisional particle model of the divertor sheath with electron emissive walls,} {\em Nuclear Fusion\/}, Vol.~63, No.~8, 2023, pp.~086022.

\bibitem{domi18}
Dom{\'\i}nguez-V{\'a}zquez, A., Cichocki, F., Merino, M., Fajardo, P., and Ahedo, E., \enquote{Axisymmetric plasma plume characterization with 2D and 3D particle codes,} {\em Plasma Sources Science and Technology\/}, Vol.~27, No.~10, 2018, pp.~104009.

\end{thebibliography}

\end{document}